\documentclass{article}
\usepackage{graphicx} % Required for inserting images
\usepackage[sort]{cite}
\usepackage[hidelinks]{hyperref}
\usepackage{booktabs}
\usepackage{spconfa4,amsmath,graphicx, acronym}
\usepackage{flushend}

\usepackage{tikz}
\usetikzlibrary{calc, positioning, fit}

\newcommand{\cmt}[1]{\textcolor{black}{#1}}

\acrodef{DNN}[DNN]{Deep Neural Network}
\acrodef{RNN}[RNN]{Recurrent Neural Network}
\acrodef{GRU}[GRU]{Gated Recurrent Unit}
\acrodef{NLR}[NLR]{Nonlinear Refinement}
\acrodef{STFT}[STFT]{Short-Time Fourier Transform}
\acrodef{SDR}[SDR]{Signal-to-Distortion power Ratio}
\acrodef{DNS}[DNS]{Deep Noise Suppression}
\acrodef{SNR}[SNR]{Signal-to-Noise power Ratio}
\acrodef{SIR}[SIR]{Signal-to-Interferer power Ratio}
\acrodef{PESQ}[PESQ]{Perceptual Evaluation of Speech Quality}
\acrodef{DnR}[DnR]{Divide-and-Remaster}
\acrodef{STOI}[STOI]{Short-Time Objective Intelligibility measure}
\acrodef{SI}[SI]{Scale-Invariant}
\title{ConcateNet: Dialogue separation Using Local and Global \\ Feature Concatenation}
\name{Mhd Modar Halimeh$^1$, Matteo Torcoli$^1$, Emanuël A. P. Habets$^2$}
\address{$^1$ Fraunhofer Institute for Integrated Circuits IIS, Erlangen, Germany\\
$^2$ International Audio Laboratories Erlangen\textsuperscript{$\ast$}, Am Wolfsmantel 33, 91058 Erlangen, Germany\thanks{\textsuperscript{$\ast$}A joint institution of Fraunhofer IIS and Friedrich-Alexander-Universit{\"a}t Erlangen-N{\"u}rnberg (FAU), Germany.}\\
{\fontfamily{qcr}\selectfont
mhd.modar.halimeh@iis.fraunhofer.de
}}

\ninept

\date{April 2024}

\begin{document}

\sloppy

\maketitle

\begin{abstract}
    Dialogue separation involves isolating a dialogue signal from a mixture, such as a movie or a TV program. This can be a necessary step to enable dialogue enhancement for broadcast-related applications. In this paper, ConcateNet for dialogue separation is proposed, which is based on a novel approach for processing local and global features aimed at better generalization for out-of-domain signals. ConcateNet is trained using a noise reduction-focused, publicly available dataset and evaluated using three datasets: two noise reduction-focused datasets (in-domain), which show competitive performance for ConcateNet, and a broadcast-focused dataset (out-of-domain), which verifies the better generalization performance for the proposed architecture compared to considered state-of-the-art noise-reduction methods. 
\end{abstract}

\begin{keywords}
    Dialogue Separation, Dialogue Enhancement, Speech Enhancement
\end{keywords}

%%%%%%%%%%%%%%%%%%%%%%%%%%%%%%%%%%%%%%%%
%%%%%%%%%%%%% Introduction %%%%%%%%%%%%%
%%%%%%%%%%%%%%%%%%%%%%%%%%%%%%%%%%%%%%%%
%\vspace*{-3mm}
\section{Introduction}
Viewers often report that dialogues in movies and other broadcast content are difficult to understand due to high background noise levels \cite{Armstrong16, Torcoli21dplus}. This is especially relevant for users who suffer from hearing impairment \cite{Shirley17}.
%, acoustic enclosures with challenging acoustic conditions, and sub-optimum reproduction devices \cite{Mapp}. %
This motivated the development of dialogue separation systems to isolate dialogue from the mixture. These systems enable re-mixing at an increased dialogue-to-background ratio, enhancing dialogue functionalities for the end-user \cite{paulus2019source}.
Dialogue separation is closely related to the general problem of speech enhancement. In particular, great overlap can be seen between dialogue separation and noise reduction, wherein both aim to extract speech signal components from a noisy mixture. 
It is also worth noting that based on the adopted definition, dialogue separation can be viewed as a special case of the more general cinematic separation problem (denoted cocktail fork problem in \cite{Cocktail}), which aims to decompose an audio mixture into, e.g., dialogue, background music, and effects components. 

Driven by the need for noise-robust telecommunication systems, noise reduction has attracted great attention in the last decades. %
%Noise reduction has attracted great attention in the last decades, especially when considering the significant performance leap achieved by utilizing \acp{DNN}. 
For instance, in \cite{Valin18}, a hybrid approach is proposed, combining conventional signal processing and \acp{DNN} to realize a low complexity, masking-based noise reduction. Alternatively, the authors in \cite{Hu2020_DCRNN} propose a complex-valued \ac{DNN}, which estimates a complex-valued mask that extracts the speech signal when applied to the noisy mixture. Element-wise multiplicative masking is extended in \cite{Mack20} such that each time-frequency bin in the extracted speech signal is obtained by filtering a corresponding region in the noisy mixture's spectrogram. A \ac{DNN} architecture is proposed in \cite{Hao21} that comprises a set of sequentially connected full- and sub-band subnetworks.  Similarly, the authors in \cite{GAGnet} propose a sequential structure using a set of modules to estimate the magnitude and complex-valued signal components iteratively. The proposed method in \cite{schroeter2023deep_mf} combines hand-crafted and learned features, which are then utilized by an autoencoder structure to extract the speech signal. An autoencoder architecture is also proposed in \cite{Rong2024}, which includes a grouped \ac{RNN} in the bottleneck layer. Reflecting their increased popularity in other domains, attention mechanisms have also been frequently employed for noise reduction, as seen in, e.g., \cite{MTFAANET,Giri19}. Notably, other proposals include architectures that are specifically designed for dialogue separation as in, e.g., \cite{Trc23, Bandsplit}. 

Nevertheless, despite the great overlap between dialogue separation and noise reduction, it is important to emphasize application-related challenges that separate the two. Namely, as dialogue separation is targeted towards broadcast-related applications, speech/dialogue signals observed by a dialogue separation system are often more diverse and can include emotional speech signals, whispering, distorted (for artistic purposes) dialogues, \cmt{overlapping speakers}, and other human sounds such as crying and screaming.
This is in contrast to publicly available noise reduction-focused datasets, which comprise mostly conversation-like, often monotone, speech utterances found in, e.g., \cite{dubey2023icassp}. This limitation holds also for publicly-available cinematic content-focused public datasets, e.g., the \ac{DnR} dataset \cite{Cocktail}, which utilizes \cite{Panayotov2015} for the speech signals. Indeed, this lack of publicly available datasets that represent realistic broadcast signals has motivated the development of a private dataset in \cite{SoundDemix} as part of the sound demixing challenge. This renders it necessary for dialogue separation systems to generalize well beyond the training conditions. Additionally, while noise reduction methods are often constrained by computational efficiency and algorithmic delay requirements, dialogue separation methods are frequently not as constrained, especially when deployed as offline systems on high-resource platforms.

\cmt{In this paper, the aim is to develop a dialogue separation approach that is robust to the mismatch between publicly-available noise reduction-focused datasets and the diverse broadcast signals observed in practice.} To this end, a \ac{DNN}-based dialogue separation method is proposed, based on separating local and global features. Consequently, this renders the proposed network capable of learning purely local, global, or hybrid patterns. As shown by the experimental results, the proposed network achieves competitive performance for conventional noise reduction datasets while outperforming considered alternatives when evaluated for a broadcast dataset.

\begin{figure*}[!t]
\centering
    \begin{tikzpicture}[]
    \node[rectangle, draw, thick, minimum width=1cm] at (0,0) (inmodule) {{Input Module}}; 
    \node[ left = 1 of inmodule] (input) {{$\mathbf{Y}$}};%(\tau)
    \node[rectangle, draw, thick, minimum width=1cm, right = 0.5 of inmodule] (FB) {{Filterbank}};
    \node[rectangle, draw, thick, minimum width=1cm, right = 0.5 of FB, fill=orange!10] (Encoder) {{Encoder}};
    \node[rectangle, draw, thick, minimum width=1cm, right = 0.5 of Encoder, fill=red!10] (TRNN) {{T-Parallel Module}};
    \node[rectangle, draw, thick, minimum width=1cm, right = 0.5 of TRNN, fill=orange!10] (Decoder) {{Decoder}};
    \node[rectangle, draw, thick, minimum width=1cm, right = 0.5 of Decoder] (MaskEst) {{Output}};
    \node[circle, draw, inner sep=0.01mm, right = 0.5 of MaskEst] (multi) {{$\times$}};
    \node[circle, draw, inner sep=0.01mm, right = 1 of multi] (add) {{$+$}};
    \node[rectangle, draw, thick, minimum width=1cm, below = 0.5 of add] (NLR) {{NLR}};
    \node[right = 0.5 of add] (output) {{$\hat{\mathbf{S}}$}};%(\tau)

    \draw[-latex, thick] (inmodule) -- (FB);
    \draw[-latex, thick] (input) -- (inmodule);
    \draw[-latex, thick] (FB) -- (Encoder);
    \draw[-latex, thick] (Encoder) -- (TRNN);
    \draw[-latex, thick] (TRNN) -- (Decoder);
    \draw[-latex, thick] (Decoder) -- (MaskEst);
    \draw[-latex, thick] (MaskEst) -- (multi);
    \draw[-latex, thick] (multi) -- (add);
    \draw[-latex, thick] (add) -- (output);
    \draw[-latex, thick] (input) -- ($(input) + (0, -0.9)$) -| (multi);
    \draw[-latex, thick] (multi) -- ($(multi) + (0.3, 0)$) |- (NLR);
    \draw[-latex, thick] (NLR) -- (add);

    \node[rectangle, minimum width=2.2cm] at ($(input) + (0.3, -1.8)$) (Encoder2) {{Encoder Module}};
    \node[rectangle, draw, thick, minimum width=2.8cm, below = 0.2 of Encoder2] (dcnn1) {{Donwsample CNN }};
    \node[rectangle, draw, thick, minimum width=2.8cm, below = 0.1 of dcnn1, fill=green!10] (dp) {{F-parallel module}};
    \node[rectangle, draw, thick, minimum width=2.8cm, below = 0.1 of dp] (dcnn2) {{CNN Module}};
    \node[rectangle, draw, thick, minimum width=2.8cm, below = 0.1 of dcnn2, rotate=0] (dcnn3) {{CNN Module}};

%    \node[rectangle, draw, thick, minimum width=2.2cm, below = 2.5 of input, rotate=90, anchor=center, fill=orange!10] (Encoder2) {\tiny{Encoder Module}};
%    \node[rectangle, draw, thick, minimum width=2.2cm, right = 0.8 of Encoder2.west, rotate=90] (dcnn1) {\tiny{Donwsample CNN }};
%    \node[rectangle, draw, thick, minimum width=2.2cm, right = 0.5 of dcnn1.west, rotate=90] (dp) {\tiny{Dual-path module}};
%    \node[rectangle, draw, thick, minimum width=2.2cm, right = 0.5 of dp.west, rotate=90] (dcnn2) {\tiny{d-wise CNN}};
%    \node[rectangle, draw, thick, minimum width=2.2cm, right = 0.5 of dcnn2.west, rotate=90] (dcnn3) {\tiny{d-wise CNN}};

%\tiny{T-RNN Module}

    \node[rectangle, right = 3 of Encoder2,  minimum width=2.2cm,] (TRNN) {{F-Parallel Module}};%{{T-RNN Module}};
    \node[coordinate] (rnninput) at ($(TRNN) - (2.2, 1.5)$) {};
    \node[left = 0.4 of rnninput, coordinate] (rnninput2) {};
    \node[rectangle, thick, draw, minimum width = 1cm] at ($(rnninput) + (1.5, 0.5)$) (rnnconv) {{CNN Module}};
    \node[rectangle, thick, draw, minimum width = 1cm, right = 0.3 of rnnconv] (gru1) {{F-GRU}};%{{T-GRU}};
    \node[rectangle, thick, draw, minimum width = 1cm] at ($(rnninput) + (1.5, -0.5)$) (skipcon) {{CNN Module}};
    \node[rectangle, thick, draw, minimum width = 1cm, right = 4 of rnninput] (cat) {{concat}};
    \node[coordinate, right = 0.3 of cat] (rnnout) {};
    
    \draw[-latex, thick] (rnninput2) -- (rnninput) |- (rnnconv);
    \draw[-latex, thick] (rnninput2) -- (rnninput) |- (skipcon);
    \draw[-latex, thick] (rnnconv) -- (gru1);
    \draw[-latex, thick] (gru1) -| (cat);
    \draw[-latex, thick] (skipcon) -| (cat);
    \draw[-latex, thick](cat)--(rnnout);

    \node[minimum width=2.2cm, right = 4.5 of TRNN] (FRNN) {{T-Parallel Module}};%{{Dual-path Module}};
    \node[coordinate] at ($(FRNN) - (2.2, 1.5)$) (rnninput) {};
    \node[left = 0.4 of rnninput, coordinate] (rnninput2) {};
    \node[rectangle, thick, draw, minimum width = 1cm] at ($(rnninput) + (1.5, 0.5)$) (rnnconv) {{CNN Module}};
    \node[rectangle, thick, draw, minimum width = 1cm, right = 0.3 of rnnconv] (gru1)  {{T-GRU}};%{{F-GRU}};
    \node[rectangle, thick, draw, minimum width = 1cm] at ($(rnninput) + (1.5, -0.5)$) (skipcon) {{CNN Module}};
    \node[rectangle, thick, draw, minimum width = 1cm, right = 4 of rnninput] (cat) {{concat}};
    \node[coordinate, right = 0.3 of cat] (frnnout) {};
    
    \draw[-latex, thick] (rnninput2) -- (rnninput) |- (rnnconv);
    \draw[-latex, thick] (rnninput2) -- (rnninput) |- (skipcon);
    \draw[-latex, thick] (rnnconv) -- (gru1);
    \draw[-latex, thick] (gru1) -| (cat);
    \draw[-latex, thick] (skipcon) -| (cat);
    \draw[-latex, thick](cat)--(frnnout);

    \node[coordinate] at ($(TRNN) + (-2.6, 0.2)$)  (fit1) {}; 
    \node[coordinate] at ($(rnnout) + (0, -1.2)$)  (fit2) {};

    \node[coordinate] at ($(FRNN) + (-2.6, 0.2)$)  (fit3) {}; 
    \node[coordinate] at ($(frnnout) + (0, -1.2)$)  (fit4) {}; 
    \node[rectangle, draw, thick, color=orange, opacity=0.3, fit = (Encoder2) (dcnn3)]{};
    \node[rectangle, draw, thick, color=green, opacity=0.3, fit = (fit1) (fit2)]{};%red
    \node[rectangle, draw, thick, color=red, opacity=0.3, fit = (fit3) (fit4)]{};%green

\end{tikzpicture}
    
    \caption{The proposed ConcateNet architecture.} \label{fig:network}
\end{figure*}
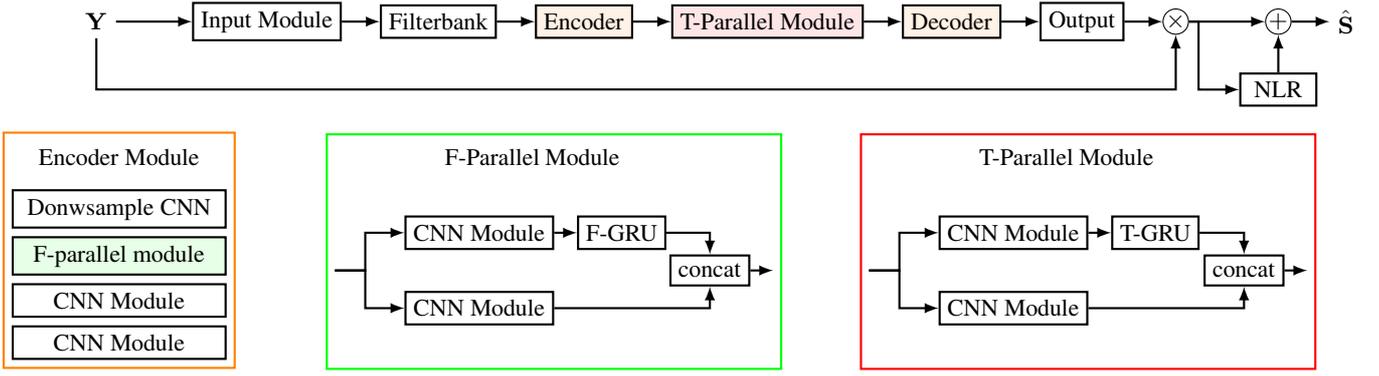

%%%%%%%%%%%%%%%%%%%%%%%%%%%%%%%%%%%%%%%%%%%%%%%
%%%%%%%%%%%%% Problem Formulation %%%%%%%%%%%%%
%%%%%%%%%%%%%%%%%%%%%%%%%%%%%%%%%%%%%%%%%%%%%%%
\section{Problem Formulation} \label{sec:problem_form}
A mixture signal $Y(m, k)$ at the time-frequency bin $(m, k)$ is modeled as
\begin{equation}
    Y(m, k) = S(m, k) + V(m ,k),
\end{equation}
where $S(m, k)$ denotes the dialogue signal, whereas $V(m, k)$ denotes the background signal. The dialogue signal contains only speech signal components \cmt{(which may involve multiple, overlapping speakers)}, while the background signal includes all other signal components, such as music, environmental noises, and sound effects. 

The problem of dialogue separation describes the task of extracting the dialogue signal $S(m, k)$ given the mixture $Y(m, k)$. This is typically achieved by estimating a complex-valued mask $M(m, k)$, which subsequently estimates the dialogue signal as
\begin{equation}
    \hat{S}(m, k) = M(m, k) Y(m, k). 
\end{equation}

Other proposals include generalizing the element-wise multiplication to multi-frame and multi-frequency-bins masking such that
\begin{equation}
    \hat{S}(m, k) =  \sum_{p=-P_1}^{P_2}\sum_{q=-Q_1}^{Q_2} M_{p, q}(m, k) Y(m-p, k-q),  \label{eq:DF}
\end{equation}
where ${P_1, P_2}$ and ${Q_1, Q_2}$ are constants that determine the spectral and temporal support and can be used to realize causal as well as non-causal filtering \cite{Mack20}. Other methods rely on directly mapping the estimated speech signal by a \ac{DNN} as in, e.g., \cite{TFGRIDNET}.

Notably, the adopted definition of dialogue separation focuses on reducing the background signal and does not include tasks such as dereverberation.

%%%%%%%%%%%%%%%%%%%%%%%%%%%%%%%%%%%%%%%%%%%
%%%%%%%%%%%%% Proposed Method %%%%%%%%%%%%%
%%%%%%%%%%%%%%%%%%%%%%%%%%%%%%%%%%%%%%%%%%%
%\vspace*{-2mm}
\section{Proposed Method} \label{sec:archiecture}
In this paper, the dialogue signal is estimated by means of a two-stage process. First, an element-wise multiplicative complex-valued mask is applied, i.e., 
\begin{equation}
    \hat{S}_1(m, k) = M(m, k) Y(m, k). \label{eq:complexmasking}
\end{equation}
Next, a \ac{NLR} step is applied such that 
\begin{multline}
    \hat{S}(m ,k) = \hat{S}_1(m ,k) \\ + \textrm{NLR}( \hat{S}_1(m + P_1, k + Q_1), \dots, \hat{S}_1(m - P_2, k - Q_2) ),  \label{eq:nonlinear_refine}
\end{multline}
where $\textrm{NLR}(\cdot)$ represents the \ac{NLR} function. The specific realization of $\textrm{NLR}(\cdot)$ is discussed in Sec.~\ref{sec:NLR}. Eq.~\eqref{eq:nonlinear_refine} can be seen as a generalization of Eq.~\eqref{eq:DF}, where the linear function (as represented by the masking operation) is replaced by a nonlinear function with an identical support range. 

The proposed architecture (denoted ConcateNet) is depicted in Fig.~\ref{fig:network}. The input to the network is the time-frequency representation of the mixture signal $\mathbf{Y}$, which is obtained by concatenating the real and imaginary components of the mixture signal along the channel axis. 

\vspace*{-2mm}
\subsection{Input Module}
The two-channel representation of the mixture signal is then processed by the \textit{input module}, which is realized as a convolutional layer, a batch normalization layer, and a ReLU activation function. This module maps the two-channel input to a higher number of channels $C$ that is maintained throughout the network. 

The output of the \textit{input module} is then processed by a Gammatone \textit{filterbank} to render $B$-bands $C$-channels spectra. Non-uniform filterbanks are also used as feature extractors in other noise reduction methods, such as \cite{schroeter2023deep_mf,MTFAANET}. 

\vspace*{-2mm}
\subsection{Encoder}
Following the input module and the filterbank, an encoder-bottleneck-decoder structure is used, where the encoder comprises three \textit{encoder modules}. Each module (shown in the orange box in Fig.~\ref{fig:network}) includes a convolutional layer that downsamples the input (along the frequency axis) using a stride of $2$, an \textit{\cmt{F-parallel} module} followed by two \cmt{convolutional} modules each comprising a convolutional layer, batch normalization, and a ReLU activation. 

The \textit{\cmt{F-parallel} module} is depicted in Fig.~\ref{fig:network} (within the green box). This module follows a two-branch structure, a \cmt{narrowband/local} branch (the lower one in Fig.~\ref{fig:network}) and a \cmt{broadband/global} branch (upper one in Fig.~\ref{fig:network}). The local branch includes a single convolutional layer, batch normalization and ReLU activation, where the convolutional layer reduces the number of input channels to $C/2$. The global branch also includes a convolutional layer, batch normalization, and ReLU activation, which reduces the number of input channels $C$ to $C/2$. This is followed by a bi-directional \ac{GRU} operating over the frequency-axis as a sequence axis (denoted F-\ac{GRU}) and uses the channels as features. Consequently, this branch processes and extracts global features from the input. Finally, the outputs of both branches are concatenated along the channel axis to recover a feature map with $C$ channels. 

\subsection{T-Parallel Module}
After the encoder, the feature map is processed at the bottleneck by the \textit{T-parallel} module. This module is shown in the red box in Fig.~\ref{fig:network}. It follows a two-branch structure similar to the \textit{\cmt{F-parallel} module} in the encoder. However, unlike the \textit{\cmt{F-parallel} module} the \textit{\cmt{T-parallel} module} utilizes a (uni-directional) T-\ac{GRU} instead of an F-\ac{GRU}. The T-\ac{GRU} operates over the time axis as a sequence dimension while utilizing the frequency axis as a feature dimension. Consequently, it is capable of exploiting temporal patterns of global feature maps. Finally, the output of the two branches is concatenated along the channel axis to recover the input's number of channels $C$.  

\subsection{Decoder}
The \textit{decoder} in Fig.~\ref{fig:network} is similarly structured to the encoder and comprises three decoder modules. Each module, in turn, comprises an upsampling convolutional layer, an \textit{\cmt{F-parallel module}}, and two \cmt{convolutional} modules each comprising a convolutional layer, batch normalization, and a ReLU activation. Upsampling is achieved using a transposed convolutional layer with a stride of $2$. As a consequence, the decoder yields an output feature map with an identical shape to that at the input of the encoder.

\subsection{\cmt{Relation to Previous Works}}
It is worth comparing ConcateNet approach of utilizing local and global features to other alternatives from the literature. The concept of extracting local and global features (also denoted frequently as sub- and full-band features) is proposed frequently, e.g., in \cite{Hao21, MTFAANET,TFGRIDNET, Bandsplit}. This emphasis on local and global features can be motivated by the ability of local (narrowband) features to generalize as they result in simple signal models. In contrast, global (broadband) features result in sophisticated signal models that can outperform their local counterparts for in-domain scenarios. 

Nevertheless, alternative proposals rely on sequentially processing local and global features, resulting in either a local map of global features or vice versa. In contrast, ConcateNet relies on separate \cmt{parallel} paths for global and local features, which allows the network to learn either purely local, purely global, or hybrid dependencies between the output and input signals. Thus, ConcateNet promises better generalization for out-of-domain recordings by preserving purely local features. 

\subsection{Mask Estimation and Nonlinear Refinement (NLR)} \label{sec:NLR}
The output of the decoder is processed by an \textit{output module}, which includes a synthesis filterbank that recovers the original number of frequency bins as the input spectra. Afterward, the synthesis filterbank is followed by a convolutional layer and a hyperbolic tangent activation function, which maps the $C$ channels to $2$ channels representing the complex-valued mask. 

Then, the \textit{initial} dialogue estimate is obtained by multiplying the input mixture signal by the complex-valued mask as described by Eq.~\eqref{eq:complexmasking}.

The \textit{initial} dialogue estimate $\hat{S}_1(m, k)$ is then processed by a \ac{NLR} module (Eq.~\eqref{eq:nonlinear_refine}) rendering the final dialogue signal estimate $\hat{S}(m, k)$. The \ac{NLR} module is constructed as a sequence of five convolutional layers (each followed by a batch normalization layer and a ReLU activation function). The first convolutional layer maps the $2$ input channels to $8$ channels, which are maintained by the subsequent layers. Finally, an additional convolutional layer is utilized to map the $8$ channels to $2$ output channels that act as real and imaginary signal components, which are added to the initial signal estimate in Eq.~\eqref{eq:nonlinear_refine}.

%%%%%%%%%%%%%%%%%%%%%%%%%%%%%%%%%%%%%%%%%%%
%%%%%%%%%% Experimental Results %%%%%%%%%%%
%%%%%%%%%%%%%%%%%%%%%%%%%%%%%%%%%%%%%%%%%%%

\section{Experimental Results} \label{sec:experimental_eval}
In the following, ConcateNet is evaluated and compared to alternative state-of-the-art methods for noise reduction. In these experiments, ConcateNet operates on time-frequency domain signals sampled at $48$~kHz obtained using \ac{STFT} with a Hamming window of $2048$ samples overlapping by $1024$ samples. Additionally, the network, including the input module, is implemented using causal depth-wise convolutional layers with $C=64$ and kernel sizes of (3,3). The filterbank within the network is realized as a Gammatone filterbank with $B=256$. The different \acp{GRU} are configured such that their output size is equal to that of their inputs. As a consequence, the number of parameters in ConcateNet is approximately $2$~M parameters.

The network is trained using the \ac{SI}-\ac{SDR} as a loss function and Adam optimizer \cite{kingma2017adam} with a step size of $0.001$. The network is trained using the \ac{DNS} challenge training dataset where the speech and noise signals are superimposed at a uniformly sampled \ac{SNR} within the range $[-5, 15]$~dB. 

Finally, four performance metrics are used to evaluate the considered methods, namely, \ac{SI}-\ac{SDR}, \ac{SI}-\ac{SIR}, \ac{PESQ} \cite{pesq}, \ac{STOI} \cite{stoi}, the 2f model \cite{fmodel}.

Three test sets are used to evaluate ConcateNet. Namely, a \ac{DNS} challenge-based test set \cite{dubey2023icassp}, the VoiceBank+Demand test set \cite{demandts}, and a private broadcast-based dataset. The comparison spans two alternative architectures, designed originally for the noise reduction task, trained identically to ConcateNet. 

\subsection{\ac{DNS} Challenge-based Test Dataset} \label{sec:exp1_dns_test}
The first test set is derived from the \ac{DNS}-challenge training dataset while ensuring that these signals are not included in the training of the different models. Specifically, two hours of speech and noise signals are mixed at randomly sampled \ac{SNR} levels within the range $[-5, 15]$~dB. Additionally, additive white Gaussian noise is added to the mixture at an \ac{SNR} level of $20$~dB. The noise and speech utterances were not included in the training dataset. 

\begin{table*}[]
    \centering
     \caption{Average performance metrics ($\pm$ standard deviation) for the different test set as described in Sec.~\ref{sec:experimental_eval}. All models were trained using the DNS challenge training dataset.}
    \begin{tabular}{l c c c c c}
    \toprule
    & SI-SDR [dB] & SI-SIR [dB] & PESQ & STOI & 2f model \\
    \midrule \\[-8pt] 
    \multicolumn{1}{l}{I. DNS Challenge-Based Test Set}\\
    \midrule
    Mixture                                                     & 5.2 ($\pm 10.02$)   &  5.1 ($\pm 10.01$)  & 1.3 ($\pm 0.46$)  & 0.45 ($\pm 0.232$)&  16.4 ($\pm 12.7$)    \\ 
    GAGNet  \cite{GAGnet}                                       & 14.4 ($\pm 6.76$)   &  26.7  ($\pm 6.75$) & 2.0 ($\pm 0.37$) & 0.57 ($\pm 0.194$)  & 36.5   ($\pm 12.0$)    \\
    DFNet 3  \cite{schroeter2023deep_mf}                        & 13.9 ($\pm 6.58$)   &  26.1 ($\pm 6.69$)  & \textbf{2.3} ($\pm 0.39$) & 0.59 ($\pm 0.198$)& 38.1  ($\pm 10.9$)\\
    ConcateNet                                                  & \textbf{15.9}  ($\pm 6.83$)  & \textbf{27.8}  ($\pm 6.17$)  & 2.1 ($\pm 0.34$) & \textbf{0.61} ($\pm 0.190$)& \textbf{39.6} ($\pm 11.9$)\\ 
    ConcateNet -/NLR                                            & 15.8 ($\pm 6.67$)   &  26.6 ($\pm 6.46$)  & 2.2 ($\pm 0.34$) &  \textbf{0.61} ($\pm 0.189$)  &  38.9 ($\pm 11.6$)\\ 
    %%%%%%%%%%%%%%%%%%%%%%%%%%%%%%%%%%%%%%%%%%%%%%%%%%%%%%%%%%%%%%%%%%%%%%%%%%%%%%%%%%%%%%%%%%%%%%%%%%%%%%%%%%%%%%%%%%%%%%
    \midrule \\[-8pt] 
    \multicolumn{1}{l}{II. VoiceBank+Demand Test Set}\\
    \midrule
    Mixture                                                     & 8.3  ($\pm 5.60$)     &  8.4 ($\pm 5.61$)     & 1.9 ($\pm 0.75$)  & 0.49 ($\pm 0.128$) & 29.3 ($\pm 15.3$)     \\ 
    GAGNet                                                      & 18.4 ($\pm 3.58$)  &  30.3 ($\pm 3.78$) & {2.9} ($\pm 0.35$)   & \textbf{0.58} ($\pm 0.129$) & 51.1 ($\pm 14.2$)     \\
    DFNet 3                                                     & 19.0 ($\pm 4.23$)  &  29.5 ($\pm 4.35$) & \textbf{3.1} ($\pm 0.36$)  & 0.56 ($\pm 0.125$) & 51.0 ($\pm 13.9$)\\
    ConcateNet                                                  & \textbf{19.5} ($\pm 4.33$)  &  25.5  ($\pm 4.45$)& 2.7 ($\pm 0.33$)  & {0.57} ($\pm 0.127$) & \textbf{51.6} ($\pm 13.6$)\\ 
    ConcateNet -/NLR                                            & 19.0 ($\pm 3.75$)  & \textbf{33.3} ($\pm 5.65$)  & 2.5 ($\pm 0.62$) & 0.56 ($\pm 0.133$) & 49.9 ($\pm 12.4$) \\ 

    \midrule \\[-8pt] 
    \multicolumn{1}{l}{III. Broadcast Test Set}\\
    \midrule
    Mixture                                                     & 7.5 ($\pm 10.9$)     &  7.7  ($\pm 10.5$) & 1.6 ($\pm 0.61$) & 0.50 ($\pm 0.207$) & 30.5 ($\pm 15.7$)    \\ 
    GAGNet                                                      & 11.7 ($\pm 7.8$)  &  23.4 ($\pm 8.0$) & 2.0 ($\pm 0.65$) & 0.59 ($\pm 0.189$) & 41.3  ($\pm 15.7$)    \\
    DFNet 3                                                     & 13.2 ($\pm 8.6$) &  25.3  ($\pm 9.4$) & \textbf{2.4} ($\pm 0.79$) &  0.61 ($\pm 0.187$) & 42.5  ($\pm 14.3$)\\
    ConcateNet                                                  & \textbf{14.2} ($\pm 8.8$) &  \textbf{26.2}  ($\pm 9.6$) & \textbf{2.4} ($\pm 0.74$) & \textbf{0.63} ($\pm 0.188$)  & \textbf{46.3} ($\pm 16.4$)\\ 
    ConcateNet -/NLR                                            & 13.7 ($\pm 8.7$) &  25.8 ($\pm 9.7$)  & 2.3 ($\pm 0.74$) & 0.62 ($\pm 0.211$) & 44.4 ($\pm 17.1$) \\ 
    \bottomrule
    \end{tabular}
   
    \label{tab:exp_results}
\end{table*}

The evaluation results are provided in Table~\ref{tab:exp_results}.I, showing that ConcateNet outperforms GAGNet \cite{GAGnet} and Deepfilternet 3 (DFNet~3)~\cite{schroeter2023deep_mf} in terms of \ac{SI}-\ac{SDR} and \ac{SI}-\ac{SIR}. Additionally, when measuring the quality of the extracted speech signals, DFNet~3 performs best w.r.t. \ac{PESQ} while ConcateNet is the best-performing model w.r.t. the 2f model (which correlates better than \ac{PESQ} with perceived quality \cite{trc}) in addition to being the best performer w.r.t. \ac{STOI}.

To quantify the benefits of the \ac{NLR} in Eq.~\eqref{eq:nonlinear_refine}, we also evaluate a ConcateNet without the \ac{NLR} module. As shown by the results in Table~\ref{tab:exp_results}.I, including the \ac{NLR} renders higher SI-SDR, SI-SIR, and 2f score, while rendering identical \ac{STOI} and a slightly lower \ac{PESQ}.  

\subsection{VoiceBank+Demand Test Dataset} \label{sec:exp2}
The VoiceBank+Demand test set \cite{demandts} is a widely used dataset to benchmark noise reduction methods. 

As shown in Table~\ref{tab:exp_results}.II, ConcateNet performs competitively when compared to the considered alternatives. More specifically, it performs best w.r.t. SI-\ac{SDR} and the 2f-model score. However, it underperforms both the GAGnet and the DFNet~3 when evaluated by the SI-SIR and \ac{PESQ} and underperforms GAGNet when evaluated by \ac{STOI}. Nevertheless, the results of the two noise reduction-focused datasets indicate state-of-the-art performance for ConcateNet and position it as a competitive noise reduction approach. 

As in the previous experiment, including the \ac{NLR} in ConcateNet improves the 2f model score, \ac{STOI}, and SI-\ac{SDR}. However, unlike the previous experiment, ConcateNet without the \ac{NLR} resulted in higher SI-\ac{SIR} and lower \ac{PESQ}.

\subsection{Broadcast Test Dataset} \label{sec:exp3}
To assess generalization to broadcast material, we use an internal broadcast dataset, which resembles typical deployment conditions for dialogue separation systems. This dataset contains broadcast dialogue and background signals. The dialogue and background signals are mixed at various \ac{SNR} levels ranging the range $[-10, 20]$~dB, with an average of $6.3$~dB. The dataset is one hour long and has $48$~kHz as sampling frequency. The dialogue signals comprise speech utterances spoken in several languages, including English and German, by male and female speakers. \cmt{These utterances vary in length between $10$~s and up to $3$~minutes.} Utterances spoken by children are also included, as well as shouting and whispering (\cmt{in contrast to, e.g., \cite{Cocktail}}). \cmt{Additionally, the dialogue signals include overlapping as well as alternating speakers (a scenario not seen during training by the networks).} The background signals included a large variety of signal types, e.g., babble noises, music, and impulsive noises, reflecting the diversity of backgrounds in broadcasting.

The performance metrics for this dataset are summarized in Table~\ref{tab:exp_results}.III. From these results, a persistent advantage for ConcateNet is observed. In particular, ConcateNet (with and without the \ac{NLR}) outperforms the considered alternatives w.r.t. SI-\ac{SDR} by $1$~dB, SI-\ac{SIR} by $0.9$~dB, and the 2f-model score by almost $4$~points. Meanwhile, identical \ac{PESQ} scores can be observed for both ConcateNet and DFNet~3. The larger performance gain for this dataset compared to the noise reduction-oriented datasets supports the hypothesis that underlines ConcateNet architecture. In particular, allowing the network to learn end-to-end local dependencies, in addition to global ones, is beneficial for the generalization of the network and can indeed enable better performance for out-of-domain signals.

Finally, confirming the benefits of the \ac{NLR} module, ConcateNet performance is increased for all performance metrics when the \ac{NLR} is utilized. This is especially true when recalling the consistent gains w.r.t. the 2f model score (which correlates strongly with the perceived signal quality) across the different test sets.

%%%%%%%%%%%%%%%%%%%%%%%%%%%%%%%%%%%%%%%%%%%
%%%%%%%%%%%%%%% Conclusion %%%%%%%%%%%%%%%%
%%%%%%%%%%%%%%%%%%%%%%%%%%%%%%%%%%%%%%%%%%%

\vspace*{4mm}
\section{Conclusion} \label{sec:conclusion}
This paper presents ConcateNet, a \ac{DNN}-based dialogue separation architecture utilizing local and global feature maps to estimate the dialogue signal. The processing of local and global feature maps is designed so that the network can preserve the local feature maps. This is motivated by the better generalization performance attributed to local features, which is a necessary characteristic for a dialogue separation system deployed for broadcast applications. In addition, an extension to multi-frame complex-valued masking is proposed by the introduction of a nonlinear refinement module at the output. As shown by the experimental results, the proposed network performs competitively when compared to alternative state-of-the-art methods for noise reduction tasks (in-domain). When evaluated for a broadcast dataset (out-of-domain), ConcateNet shows a decisive advantage over the considered baselines, underscoring the better generalization capabilities of the architecture.

\clearpage
\newpage

\bibliographystyle{IEEEbib}
\bibliography{refs}
\end{document}